\documentclass[letterpaper,aps,prl,twocolumn,amsmath,showpacs,amssymb]{revtex4-1}
\pdfoutput=1
\usepackage{color}
\usepackage[pdftex]{graphicx,hyperref}
\usepackage{dcolumn}
\usepackage{bm}
\usepackage{marvosym}
\usepackage{times,amsmath,amssymb}

\begin{document}
\title{Tunable graphene-based polarizer}

\author{Yu. V. Bludov}
\email{bludov@fisica.uminho.pt}
\author{M. I. Vasilevskiy}
\author{N. M. R. Peres}

\affiliation
{Centro de F\'{\i}sica \& Departamento de F\'{\i}sica, Universidade do Minho,
Campus de Gualtar, Braga 4710-057, Portugal
}

\begin{abstract}
It is shown that an attenuated total reflection structure containing a graphene layer
can operate as a tunable polarizer of the electromagnetic radiation. The polarization angle
is controlled by adjusting the voltage applied to graphene via external gate. The mechanism is based on the resonant coupling
of $p-$polarized electromagnetic waves to the surface plasmon-polaritons in graphene. The presented calculations show that, at resonance, the reflected wave is almost 100\% $s-$polarized.
\end{abstract}
\maketitle

Surface plasmon-polaritons (SPPs) are excitations of electromagnetic (EM) radiation coupled to surface charges existing at a metal-dielectric interface.\cite{c:maradudin,Zhang2012} Under certain conditions, they can interact with external EM waves giving rise to a number of optical effects ranging from resonant absorption and enhanced non-linearities to surface-enhanced Raman scattering.\cite{Stockman} The potential applications of SPPs in detection, signal processing and information transfer originated the appearance of a new and rapidly growing research area called plasmonics.\cite{Stockman,c:maier}

Graphene, a monolayer thick metal is attracting much attention in different areas \cite{c:geim-nov-2007,c:geim-2009,c:rev-chem} due to its unique and unusual physical properties. For plasmonics, the interest in graphene stems from the possibility of controlling the charge carrier density through electrostatic "doping" using an external gate voltage.\cite{c:cond-experiment,c:graphene-plas-tuning}
As a consequence, the spectrum of SPPs in graphene is tunable as it is determined by the optical conductivity that depends on the free carrier concentration.\cite{c:peres_RMP} This idea is supported by the observation of tunable plasmon absorption bands in graphene ribbons \cite{c:ribbon-experiment} and the demonstration of a terahertz (THz) source,\cite{c:graphene-plas-source} and opens the possibility of creation of graphene-based metamaterials and new optoelectronic devices, such as a THz detector,\cite{c:ribbon-experiment} a broad band modulator \cite{c:graphene-modulator} and an optical switch.\cite{c:my} Increasing the electron density by charge doping \cite{charge doping} can make these devices work also in the infrared domain.

The EM response of graphene is characterized by its optical conductivity that can be written as $\sigma(\omega)=\sigma'+i\sigma''$,
where $\sigma'$ and $\sigma''$ are real functions of frequency, $\omega $. A unique property of doped graphene is that
the imaginary part of the conductivity, $\sigma''$, can be either positive or negative depending on the frequency and the electron chemical potential, $\mu$. It is determined by both intraband scattering (Drude term) and interband transitions, and reaches a (negative value) minimum at $\hbar \omega \approx 2 \mu$.\cite{c:peres_RMP} As a consequence, in contrast with usual metals,
graphene can support either transverse magnetic (TM) or transverse electric (TE) surface waves, depending on whether $\sigma''$
is positive or negative.\cite{c:mikhailov} The existence of TE-type (or $s-$polarized) SPPs is unique to graphene and has been demonstrated experimentally.\cite{c:polarizer} Let us note at this point that their dispersion relation is quite different from that of "conventional" TM-type (or $p-$polarized) SPPs, as it will be further discussed below.

\begin{figure}[h!]
\includegraphics[width=8.5cm]{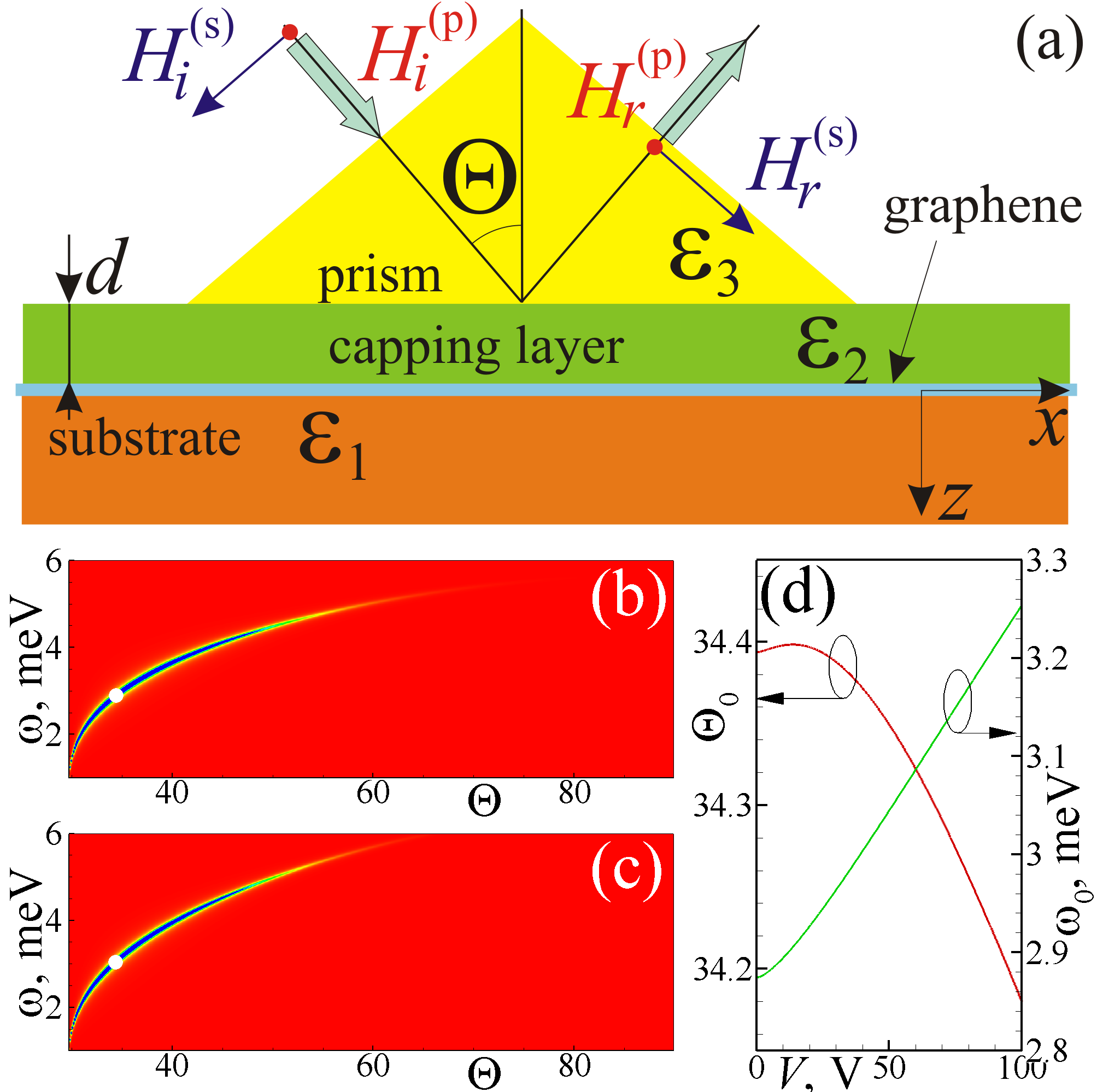}\\
\caption{(a) ATR structure in Otto configuration, containing a graphene layer
on the substrate, and a prism separated from graphene by
a clapping layer of thickness $d$; (b, c) Reflectivity $R=\left|H_{r}^{(p)}\right|^{2}$
versus angle of incidence $\Theta$, and frequency $\omega$ for the
ATR structure with $\mu\simeq0.1\,$eV (b), or $\mu\simeq0.222\,$eV
(c) {[}corresponding to $V=10\,$V and $50\,$V of gate voltage,
respectively{]}, where blue (red) color corresponds to low (high) reflectivity; (d) Angle of incidence $\Theta_{0}$, and frequency
$\omega_{0}$, corresponding to zero reflectivity of the ATR structure
versus gate voltage $V$. In all panels other parameters are $\varepsilon_{1}=3.9$,
$\varepsilon_{2}=1$, $\varepsilon_{3}=16$, $d=40\,\mu\mathrm{m}$,
$\varphi=90^{\circ}$ ($p-$polarized wave). In panels (b, c) white circles
depict pairs of parameters ($\Theta_{0},\omega_{0}$) corresponding (in each case)
to zero reflectivity.}
\label{fig:reflectivity}
\end{figure}

Conventional polarizers use one of the following operation principles: (i) polarization-dependent absorption in anisotropic media, (ii) refraction by a prism, and (iii) polarization by reflection at Brewster angle.\cite{polarizers} Recently, the idea of filtering of one of the polarizations using a graphene-based device has been proposed and it allowed for demonstration of a polarizer that can operate in the infrared range.\cite{c:polarizer} The device consisted of an optical fiber some part of which was side-polished and cladded with a graphene sheet. Due to the higher attenuation of the $p-$component of the EM field (if the radiation is in resonance with TE plasmon-polaritons), a rather high degree of the TM component filtering could be achieved, given the graphene-cladded segment of the fibre is long enough (though, the TE component is attenuated, too). We propose another possible operation principle for a graphene based polarizer, making use of the resonant coupling of the external EM radiation to SPPs in an attenuated total reflection (ATR) structure considered in our previous work \cite{c:my} and schematically shown in Fig. \ref{fig:reflectivity}.

In this letter we demonstrate the possibility of rotating the plane of polarization of incoming light by its reflection in an ATR graphene-based structure. The rotation angle can be controlled by varying the external gate voltage applied to graphene. In other words,
we will show that such a graphene-based ATR structure can operate as a tunable polarizer of the EM radiation. At the resonant frequency depending on the gate voltage, the $p-$component of the incident wave is almost 100\% filtered out.

Let us consider the ATR scheme for coupling SPPs to an external EM wave using so called Otto configuration\cite{Zhang2012} (see Fig. \ref{fig:reflectivity}). In this configuration, a
linearly polarized EM wave with frequency $\omega$ impinges on the interface between the prism, with a dielectric constant
$\varepsilon_{3}$, and a graphene layer sandwiched between a substrate (with a dielectric constant $\varepsilon_{1}$) and a capping layer (with a dielectric constant $\varepsilon_{2}$). We assume that all these dielectric constants are real. If
the angle of incidence $\Theta$ is larger than the critical angle of total internal reflection $\Theta_{c}$
[determined by the relation $\sin\Theta_{c}={\rm max}(\varepsilon_{1},\varepsilon_{2})/\varepsilon_{3}$],
then the EM wave in the gap between the prism and graphene
will be evanescent in the $z$ direction. The same is true for the field
in the substrate ($z>0$).

Two situations, qualitatively different in terms of coupling of the external EM wave to SPPs in graphene, are possible: (i) if the $x$-component of the incident wave vector, $k_{x}=(\omega/c)\sqrt{\varepsilon_{3}}\sin\Theta$, does not match $k_{SPP}(\omega)$, the SPP wave vector in graphene, there is virtually no interaction between them and the incident wave is almost totally reflected at the prism-capping layer interface (the reflectivity is close to unity); (ii) if $k_{x}\simeq k_{SPP}(\omega)$, the incident EM wave and the polariton mode are in resonance and the impinging energy is transferred to the excitated SPPs in graphene resulting in a drastic decrease of the reflectivity.
The operation mechanism of the polarizer is now easy to understand. Assuming that the
frequency of the impinging radiation lies in the spectral region where $\sigma''>0$, $p-$polarized SPPs can be induced in graphene. A linear polarized wave with an arbitrary polarization orientation can be represented as a sum of TE and TM waves without phase shift. If the wave is in resonance with SPP, the $p-$polarized component of the {\it reflected} wave is strongly attenuated whereas the $s-$polarized one has almost the same amplitude as in the incident wave. This leads to a rotation
of the polarization plane of the reflected EM wave with respect to that of the impinging radiation.

In order to describe the proposed mechanism quantitatively,
we solve Maxwell's equations, ${\rm rot}\mathbf{E}^{(m)}=i\kappa\mathbf{H}^{(m)}$,
${\rm rot}\mathbf{H}^{(m)}=-i\kappa\varepsilon_{m}\mathbf{E}^{(m)}$,
separately for each of three media ($m=1,2,3$). Since the system is
uniform in the $y$ direction, we can decompose the electromagnetic fields $\mathrm{\mathbf{E}}^{(m)},$
$\mathbf{H}^{(m)}$ into two components and consider the $s$ and $p-$polarized waves separately. For the latter, the magnetic field vector is perpendicular to the plane of incidence ($xz$) and we have $\mathbf{H}^{(m)}=\{0,\, H_{y}^{(m)},\,0\}$ and $\mathbf{E}^{(m)}=\{E_{x}^{(m)},\,0,\, E_{z}^{(m)}\}$. For the $s-$polarized wave, the magnetic field vector lies in the plane of incidence, $\mathbf{H}^{(m)}=\{H_{x}^{(m)},\,0,\, H_{z}^{(m)}\}$) and $\mathbf{E}^{(m)}=\{0,\, E_{y}^{(m)},\,0\}$. The amplitudes of $p$ and $s-$polarized parts of the incident wave are $H_{i}^{(p)}=H_{i}\sin\varphi$,
and $H_{i}^{(s)}=H_{i}\cos\varphi$, where $\varphi$ denotes the
angle between the plane of incidence and the direction of the magnetic field, and the phase is the same for both waves.

In the
semi-infinite medium $m=3$ (representing the prism and occupying the half-space $z<-d$) the solution of Maxwell's equations reads:
\begin{eqnarray}
&&H_{x}^{(3)}(x,z)=-\exp(ik_{x}x)\cos\Theta\times\label{eq:hx3}\\
&&\{H_{i}\exp[ik_{z}(z+d)]\cos\varphi-H_{r}^{(s)}\exp[-ik_{z}(z+d)]\},\nonumber \\
&&E_{y}^{(3)}(x,y)=\frac{1}{\sqrt{\varepsilon_{3}}\sin\Theta}H_{z}^{(3)}=\exp(ik_{x}x)\frac{1}{\sqrt{\varepsilon_{3}}}\times\label{eq:ey3}\\
&&\{H_{i}\exp[ik_{z}(z+d)]\cos\varphi+H_{r}^{(s)}\exp[-ik_{z}(z+d)]\},\nonumber
\end{eqnarray}
for the $s-$polarized wave, and
\begin{eqnarray}
&&H_{y}^{(3)}(x,z)=-\frac{\sqrt{\varepsilon_{3}}}{\sin\Theta}E_{z}^{(3)}(x,z)=\exp(ik_{x}x)\times\label{eq:hy3}\\
&&\{H_{i}\exp[ik_{z}(z+d)]\sin\varphi+H_{r}^{(p)}\exp[-ik_{z}(z+d)]\},\nonumber \\
&&E_{x}^{(3)}(x,z)=\exp(ik_{x}x)\frac{\cos\Theta}{\sqrt{\varepsilon_{3}}}\times\label{eq:ex3}\\
&&\{H_{i}\exp[ik_{z}(z+d)]\sin\varphi-H_{r}^{(p)}\exp[-ik_{z}(z+d)]\},\nonumber
\end{eqnarray}
for the $p-$polarized one. We have defined $k_{z}=\kappa\sqrt{\varepsilon_{3}}\cos\Theta$ as the $z$-component
of the wave vector, $\kappa =\omega /c$. Note that two limiting cases, $\varphi=0$ and
$\varphi=\pi/2$, correspond to purely $s$ and $p-$polarized incident waves, respectively.
In this cases, the reflected wave is linear polarized. For an arbitrary $\varphi$ the (complex) amplitudes of the $p$ and $s-$polarized reflected waves, $H_{r}^{(p)}$ and $H_{r}^{(s)}$,
do not have the same phase. In general, this phase difference gives rise to an elliptic polarization of the reflected wave.

In the dielectric film between the prism and graphene ($-d<z<0$, medium
$m=2$), and in the substrate below graphene ($z>0$, medium $m=1$) the solution
of Maxwell's equations is:

\begin{eqnarray}
&&H_{x}^{(m)}(x,z)=\exp(ik_{x}x)\times\nonumber\\
&&\left\{ A_{+}^{(m)}\exp[p_{m}z]+A_{-}^{(m)}\exp[-p_{m}z]\right\} ,\label{eq:hx2} \\
&&E_{y}^{(m)}(x,y)=\frac{1}{\sqrt{\varepsilon_{3}}\sin\Theta}H_{z}^{(m)}=-\exp(ik_{x}x)\frac{i\kappa}{p_{m}}\times\nonumber\\
&&\left\{ A_{+}^{(m)}\exp[p_{m}z]-A_{-}^{(m)}\exp[-p_{m}z]\right\} ,\label{eq:ey2}
\end{eqnarray}
for the $s-$polarized wave, and
\begin{eqnarray}
&&H_{y}^{(m)}(x,z)=-\frac{\varepsilon_{m}}{\sqrt{\varepsilon_{3}}\sin\Theta}E_{z}^{(m)}(x,z)=\exp(ik_{x}x)\times\nonumber\\
&&\left\{ B_{+}^{(m)}\exp[p_{m}z]+B_{-}^{(m)}\exp[-p_{m}z]\right\} ,\label{eq:hy2} \\
&&E_{x}^{(2)}(x,z)=-i\frac{p_{m}}{\kappa\varepsilon_{m}}\exp(ik_{x}x)\times\nonumber \\
&&\left\{ B_{+}^{(m)}\exp[p_{m}z]-B_{-}^{(m)}\exp[-p_{m}z]\right\} ,\label{eq:ex2}
\end{eqnarray}
for the $p-$polarized one. The fields are represented in the form of evanescent waves [in Eqs. (\ref{eq:hx2}) - (\ref{eq:ex2})
$p_{m}=\kappa\sqrt{\varepsilon_{3}\sin^{2}\Theta-\varepsilon_{m}}\geq0$].
The case $m=2$ corresponds to a superposition of two evanescent waves, while for $m=1$ we must have $A_{+}^{(1)}=B_{+}^{(1)}=0$, that is, the solution represents a single wave of the amplitude exponentially decreasing with the distance
$z$ to graphene.

The boundary conditions at $z=-d$ and $z=0$ are different.
At the prism-dielectric interface, $z=-d$, they imply the continuity of the tangential components of the electric
and magnetic fields, $E_{x,y}^{(3)}(x,-d)=E_{x,y}^{(2)}(x,-d)$, $H_{x,y}^{(3)}(x,-d)=H_{x,y}^{(2)}(x,-d)${]}.
At the interface $z=0$ the condition is different due to the presence
of the conductive graphene layer. In this case, we have  the
continuity of the
tangential components of the electric field, $E_{x,y}^{(1)}(x,0)=E_{x,y}^{(2)}(x,0)$, whereas
 the tangential component of the magnetic field is discontinuous, that is, we have
$H_{y}^{(1)}(x,0)-H_{y}^{(2)}(x,0)=-(4\pi/c)j_{x}=-(4\pi/c)\sigma(\omega)E_{x}(x,0)$,
$H_{x}^{(1)}(x,0)-H_{x}^{(2)}(x,0)=(4\pi/c)j_{y}=(4\pi/c)\sigma(\omega)E_{y}(x,0)$.
Matching these boundary conditions, we obtain explicit expressions
for the amplitudes of the reflected fields,
\begin{eqnarray}
  H_{r}^{(s)}&=&H_{i}\cos\varphi\frac{\left\{ p_{1}-i\frac{4\pi\kappa}{c}
\sigma(\omega)\right\} \overline{\eta_{1}}+p_{2}\overline{\eta_{2}}}
{\left\{ p_{1}-i\frac{4\pi\kappa}{c}\sigma(\omega)\right\} \eta_{1}+p_{2}\eta_{2}},\label{eq:hrs}
\end{eqnarray}
for the $s-$polarized component, and
\begin{eqnarray}
 H_{r}^{(p)}&=&-H_{i}\sin\varphi\frac{\left\{ \frac{\varepsilon_{1}}
{p_{1}}+i\frac{4\pi}{c\kappa}\sigma(\omega)\right\}
\overline{\chi_{1}}+\frac{\varepsilon_{2}}{p_{2}}
\overline{\chi_{2}}}{\left\{ \frac{\varepsilon_{1}}{p_{1}}+i\frac{4\pi}{c\kappa}
\sigma(\omega)\right\} \chi_{1}+\frac{\varepsilon_{2}}{p_{2}}\chi_{2}},\label{eq:hrp}
\end{eqnarray}
for the $p-$polarized one, where we have defined
\begin{eqnarray*}
\eta_{1}&=&\tanh[p_{2}d]+\frac{ip_{2}}{\kappa\sqrt{\varepsilon_{3}}\cos\Theta},\\ \eta_{2}&=&1+\frac{ip_{2}}{\kappa\sqrt{\varepsilon_{3}}\cos\Theta}\tanh[p_{2}d],\\
\chi_{1}&=&\tanh[p_{2}d]-\frac{i\kappa\varepsilon_{2}\cos\Theta}{p_{2}\sqrt{\varepsilon_{3}}},\\
\chi_{2}&=&1-\frac{i\kappa\varepsilon_{2}\cos\Theta}{p_{2}\sqrt{\varepsilon_{3}}}\tanh[p_{2}d],
\end{eqnarray*}
and the bars over $\eta_{1}$ and $\chi_{1/2}$ denote complex conjugates.
\begin{figure}
\includegraphics[width=8.5cm]{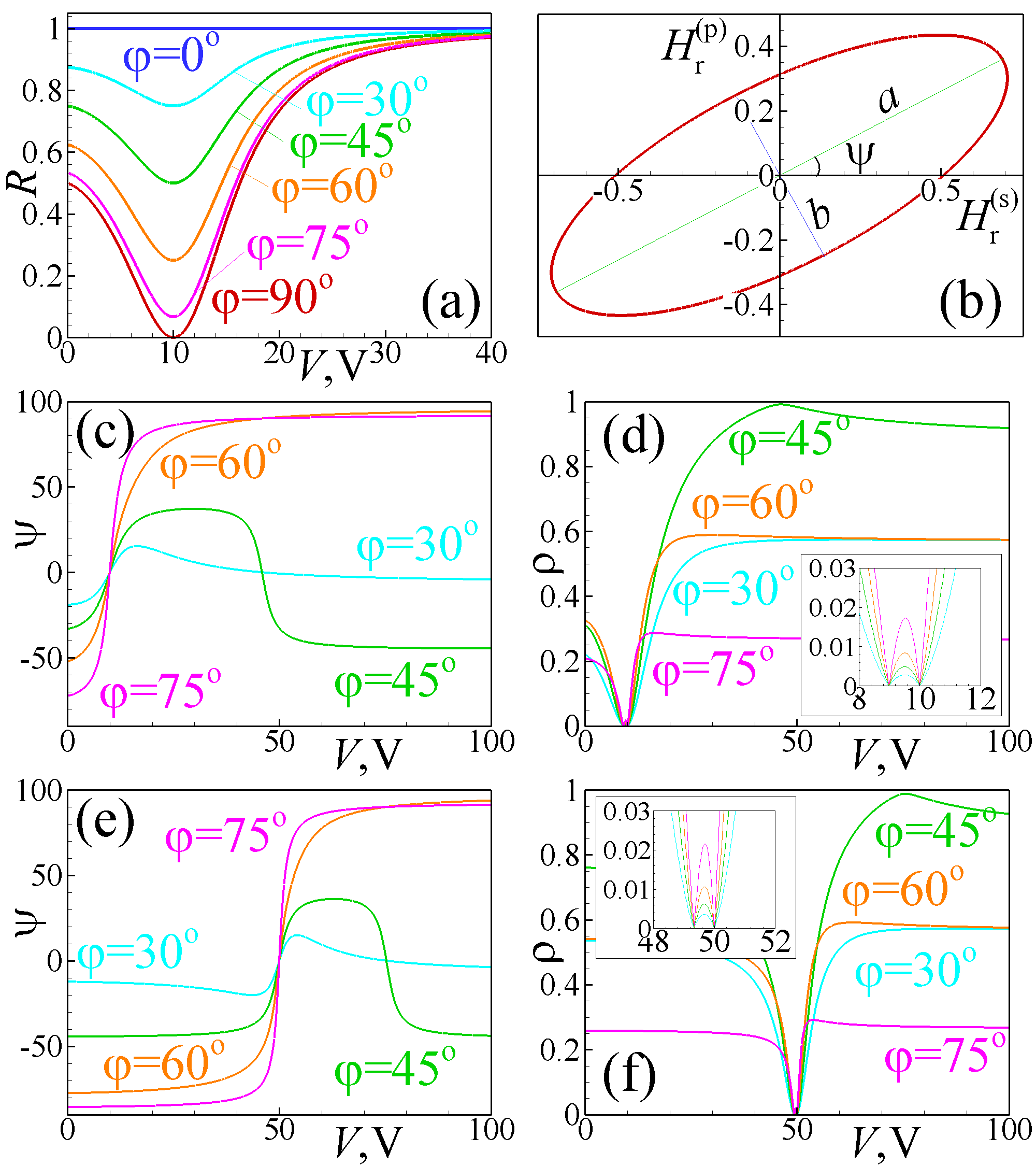}
\caption{(a) Reflectivity $R=\left|H_{r}^{(p)}\right|^{2}+\left|H_{r}^{(s)}\right|^{2}$
of EM waves with different polarization angles $\varphi$ (as indicated on the plot) {\it versus} gate voltage;
(b) Vibrational ellipse of the magnetic field vector in the reflected wave for $V=15\,\mathrm{V}$; (c, d, e, f) Dependence of the orientation angle of the ellipse, $\psi$ (c,e) and the ellipticity, $\rho$ (d, f) upon the applied gate voltage for different polarization angles of the incident wave (as indicated on the plot) and $\Theta=34.40\,^{\circ}$, $\hbar \omega=2.894\,\mathrm{meV}$ (c, d) or $\Theta=34.35\,^{\circ}$, $\hbar \omega=3.044\,\mathrm{meV}$ (e, f).
Other parameters of the ATR structure are the same as in Fig. \ref{fig:reflectivity}.
The insets in (d, f) show zoom into the region near the minimum of $\rho $.
}
\label{fig:polar}
\end{figure}

Let us analyze the properties of Eqs. (\ref{eq:hrs})--(\ref{eq:hrp}). We note
that in the limit of an infinitely thick layer separating the prism and graphene, $d\to\infty$,
the denominator of Eq. (\ref{eq:hrs}) is equal to zero if
\begin{equation}
p_{1}+p_{2}-i\frac{4\pi\kappa}{c}\sigma(\omega)=0,\label{eq:ev-s}
\end{equation}
while  for Eq. (\ref{eq:hrp}) the same condition yields,
\begin{equation}
\frac{\varepsilon_{1}}{p_{1}}+\frac{\varepsilon_{2}}{p_{2}}+i\frac{4\pi}{c\kappa}\sigma(\omega)=0.\label{eq:ev-p}
\end{equation}
Substitution of $k_{SPP}=\kappa\sqrt{\varepsilon_{3}}\sin\Theta$
into Eqs. (\ref{eq:ev-s}) and (\ref{eq:ev-p}) transforms them
into the dispersion relations, $\omega(k_{SPP})$, for TE {[}Eq. (\ref{eq:ev-s}){]},
and TM {[}Eq. (\ref{eq:ev-p}){]} surface waves in graphene cladded
by two semi-infinite media with dielectric constants $\varepsilon_{1}$ and $\varepsilon_{2}$.
Despite  some algebraic similarity between these relations, the $p$ and $s-$ polarized SPPs possess
quite different properties. As mentioned above, Eq. (\ref{eq:ev-p}) has solution, i.e. $p-$polarized SPPs exist if $\sigma''>0$,
whereas solution of Eq. (\ref{eq:ev-s}) ($s-$polarized surface wave) exists in the opposite case.
Taking into account the form of the optical conductivity of
graphene \cite{c:peres_RMP,c:cond-experiment}, we conclude that
the two types of waves exist in different spectral ranges, namely, $p-$polarized SPPs are present at low frequencies ($\hbar \omega < 2\mu$),
while $s-$polarized waves exist at frequencies in the vicinity of $\hbar \omega\sim 2\mu$, where the chemical potential depends on the gate voltage, $\mu \sim \sqrt {V}$.

Since the working principle of the polarizer is based on the excitation of TM SPPs, in Figs. \ref{fig:reflectivity}(b-d) we present the $p-$polarized reflectance spectra in detail. As discussed above, resonant
excitation of SPPs for a frequency $\omega$ is manifested by the appearance
of a reflectivity minimum {[}Figs. \ref{fig:reflectivity}(b,c){]} at
a certain angle $\Theta$. Excitation
of SPPs in graphene is characterized by two interesting properties.
First, the position of the resonant reflectivity minimum can
be tuned by changing the gate voltage applied to the graphene layer,
as it follows from the comparison of Figs. \ref{fig:reflectivity}(b)
and \ref{fig:reflectivity}(c). Secondly, if the incident wave is purely
$p-$polarized ($\varphi=90^{\circ}$), and for any given value of the gate voltage,
it is possible to find a pair of parameters, $(\Theta_{0},\omega_{0})$, for
which the reflectivity of the ATR structure is zero. In Figs. \ref{fig:reflectivity}(b,c)
these pairs are depicted by white circles. Below, the pair $(\Theta_{0},\omega_{0})$
for a given $V$
is referred to as {\it zero-point}. In other words, at the zero-point the
whole energy of the incident wave is transformed into energy of excited
SPPs. In Figs. \ref{fig:reflectivity}(d) the pair $(\Theta_{0},\omega_{0})$
is shown as a function of the gate voltage.

How should the tunable polarizer work?
If the parameters $V$ , $\Theta$, and $\omega$ are chosen
is such a way that $\Theta=\Theta_0$ and $\omega=\omega_0$ for the given $V$,
and the incident wave is not purely $p-$polarized,
($\varphi\neq90^{\circ}$), then its $p-$polarized part is completely
absorbed, while the $s-$polarized part is almost totally reflected,
giving rise to the total reflectivity $R\approx\cos^{2}\varphi$
{[}see Fig. \ref{fig:polar}(a){]}. In this case we have pure $s-$polarization of the reflected EM radiation with a single passage through the device.

Keeping unchanged $\Theta$ and $\omega$, and slightly detuning $V$ from its initial value
deviates the system from the condition necessary for the full absorption of the $p-$polarized component.
The reflected wave then should contain the fully reflected $s-$polarized component
as well as a small $p-$polarized part.
Therefore, the total reflectivity $R$ increases {[}Fig. \ref{fig:polar}(a){]}
and, in general, the reflected light will be elliptically polarized
{[}an example is shown in Fig. \ref{fig:polar}(b){]}. Further detuning
of the gate voltage will result in a further increase of the total reflectivity and the corresponding decrease of the polarization ratio.

In what follows we want to characterize the
reflected wave in more detail, namely, to investigate the shape and orientation of the vibrational ellipse characterized by
the orientation angle $\psi$ and the ellipticity $\rho=b/a$ \cite{Born-Wolf}
{[}see Fig. \ref{fig:polar}(b) for definitions of these quantities{]}
as functions of the gate voltage.
Figs. \ref{fig:polar}(c-f) summarize our results.
At the zero-point, pure $s-$polarization of the reflected wave corresponds to $\psi=0$ {[}Figs. \ref{fig:polar}(c) and
 \ref{fig:polar}(e){]} and zero ellipticity {[}Figs. \ref{fig:polar}(d)
and \ref{fig:polar}(f){]}. In its vicinity, an increase
of the gate voltage results in an anti-clockwise rotation of the vibrational ellipse, since $\psi$ increases as seen in Figs. \ref{fig:polar}(c) and \ref{fig:polar}(d).
Far from the zero-point, the increase of $\psi$ is monotonic
for  $\varphi>45^{\circ}$ and has a maximum at a certain $V$
for $\varphi\leq45^{\circ}$.
When the  gate voltage $V$ is detuned from the zero-point, the reflected wave ellipticity
$\rho$ becomes non-zero. However, as it follows from the insets
of Figs. \ref{fig:polar}(d) and \ref{fig:polar}(f), in the vicinity
of the zero point there is another value of $V$, $V\approx9\,\mathrm{V}$ in
Fig. \ref{fig:polar}(d) and $V\approx49.3\,\mathrm{V}$ in Fig. \ref{fig:polar}(f),
at which the reflected wave is linearly polarized ($\rho=0$). Although
this point bears some similarity with the zero-point, the reason for $\rho=0$ is different.
Here the linear polarization of the reflected wave stems from the
phase shift (equal to $\pi$) between the $s$ and $p-$components, while at the zero-point it is due to the full absorption
of its $p-$component.
We also notice that a steeper variation of $\psi$ with $V$, close to the zero point,
occurs for larger values of $\varphi$. The same happens for the
ellipticity $\rho$. For $\varphi\lesssim90^\circ$, $\rho $ is close to zero (linear polarization). Therefore, for these angles of polarization of the incident wave. the reflected radiation presents a large $\psi$ and a small ellipticity for $V$ slightly detuned from the zero point.

To conclude, we have demonstrated that an ATR structure containing a single graphene layer can work as a tunable polarizer of the EM radiation in the THz frequency domain, in a broad range of gate voltages.
Although we have considered the case where the incident light
is linearly polarized, it is possible to generalize our results to
the cases of e.g. circular polarized incident electromagnetic radiation. Also, we would like to point out the possibility of using the same principle for filtering out $s-$polarized component of the incident wave. It would allow to extend the operation frequency range to the infrared domain. However, our calculations show that, in general, the resonance with TE polaritons in graphene is weaker than with $p-$polarized SPPs.

{\it Acknowledgement.} The work was partially supported by
the Portuguese Foundation for Science and Technology.



\begin{thebibliography}{10}
\bibitem {c:maradudin}A. A. Maradudin, Surface Polaritons. Electromagnetic
Waves at Surfaces and Interfaces, ed V. M. Agranovich and D. L. Mills,
Amsterdam, North-Holland, 1982.

\bibitem {Zhang2012} J. Zhang, L. Zhang, and W. Xu, J. Phys. D: Appl. Phys. 45, 113001 (2012).

\bibitem {Stockman}M. I. Stockman, Phys. Today 64, 39 (2011).

\bibitem {c:maier} S. Maier, {\it Plasmonics: Fundamentals and Applications},
Springer, 2007.

\bibitem {c:geim-nov-2007}A. K. Geim and K. S. Novoselov, Nature
Materials 6, 183 (2007).

\bibitem {c:geim-2009} A. K. Geim, Science 324, 1530 (2009).

\bibitem {c:rev-chem} M. J. Allen, V. C. Tung and R. B. Kaner, Chem.
Rev. 110, 132 (2010).

\bibitem {c:cond-experiment} Z. Q. Li, E. A. Henriksen, Z. Jiang, Z. Hao, M. C. Martin, P. Kim, H. L. Stormer, and D. N. Basov, Nature Phys. 4, 532 (2008).

\bibitem {c:graphene-plas-tuning} V. Ryzhii, A. Satou, and T. Otsuji,
J. Appl. Phys. 101, 024509 (2007)

\bibitem {c:peres_RMP}
N. M. R. Peres, Rev. Mod. Phys. 82, 2673 (2010).

\bibitem {c:ribbon-experiment} L. Ju, B. Geng, J. Horng,
C. Girit, M. C. Martin, Z. Hao, H. A. Bechtel, X. Liang, A. Zettl, Y. Ron Shen, and F. Wang, Nature Nanotechnol.
6, 630 (2011).

\bibitem {c:graphene-plas-source}V. Ryzhii, M. Ryzhii, and T. Otsuji,
J. Appl. Phys. 101, 083114 (2007); F. Rana, IEEE Trans. NanoTechnol.
7, 91 (2008).

\bibitem {c:graphene-modulator} M. Liu, X. Yin, E. Ulin-Avila, B. Geng, T. Zentgraf, L. Ju, F. Wang, and X. Zhang, Nature 474, 64 (2011).

\bibitem {c:my} Yu. V. Bludov, M. I. Vasilevskiy, and N. M. R. Peres, Europhys. Lett. 92, 68001 (2010).

\bibitem {charge doping} S. Y. Shin {\it et.al,}, Appl. Phys. Lett. 99, 082110 (2011).

\bibitem {c:mikhailov} S. A. Mikhailov and K. Ziegler, Phys.
Rev. Lett. 99, 016803 (2007).

\bibitem {c:polarizer}Qiaoliang Bao, Han Zhang, Bing Wang, Zhenhua
Ni, Candy Haley Yi Xuan Lim, Yu Wang, Ding Yuan Tang, and Kian Ping
Loh, Nature Phot. 5, 411 (2011).

\bibitem {polarizers} D. Goldstein, {\it Polarized Light}, Marcel Dekker Inc., New York, 2003.

\bibitem{Born-Wolf} M. Born and E. Wolf, {\it Principles of Optics}, Pergamon, Oxford (1989).


\end{thebibliography}
\end{document}